\documentclass{article}
\usepackage{float}
\usepackage{amsmath}
\usepackage{comment}
\usepackage{graphicx} 
\usepackage[
backend=biber,
style=numeric,
sorting=ynt
]{biblatex}

\providecommand{\keywords}[1]{\textbf{\textit{Keywords:}} #1}

\title{Expected by Whom? A Skill-Adjusted Expected Goals Model for NHL Shooters and Goaltenders}
\author{J.T.P. Noel}
\date{November 2025}

\begin{document}

\maketitle

\begin{abstract}
    This study outlines a light gradient boosted model aimed at predicting shot outcomes in the NHL. The model uses the NHL's spatiotemporal data to account for both the skill of shooters and goaltenders. This approach involves isolating and engineering features for different aspects of shooter and goaltender skill. These aspects include the overall skill, the locational skill, which is engineered using a shot binning technique previously outlined by Shuckers and Curro, and the situational skill, which is engineered using Gower distance. Three separate datasets were created based on the skill of the shooter and goaltender. For each, a baseline model was created in order to compare and contrast its performance with the skill-adjusted model. The results seen in this study show performance increases for the skill-adjusted model over the baseline model in log loss, brier scores, and area under the ROC curve. These performance increases have a high of 5\% and outperform previous works, which have attempted to account only for player skill. This highlights the importance of accounting for both player and goaltender skill, while also accounting for different aspects of their skill. In future works, a skill-adjusted expected goals model could benefit models interested in predicting other aspects of the game, such as scoring leaders or individual game outcomes. 
\end{abstract}


\section{Introduction}
In recent years, analytics has changed the way teams, fans and players approach and understand the game of ice hockey \cite{nandakumar2019historical}. This new age of analytics has ushered in several new performance indicators that can improve team evaluation, player evaluation, and the understanding of what ultimately leads to a team winning or losing. One of the more popular performance indicators to come from this era is expected goals (xG) \cite{johansson2022analytics}. Expected goals is a measurement of shot quality that typically uses spatiotemporal data and machine learning (ML) to assign a likelihood that a given shot attempt will result in a goal.

Hockey's introduction to expected goals is often attributed to Alan Ryder, who in 2004 proposed a method to account for shot quality in the NHL. Ryder's work opens with a phrase that has become synonymous with expected goals ``Not all shots on goal are created equal''\cite{ryder}. This phrase perfectly encapsulates the concept behind expected goals, rather than looking at how many shots a given team or player has taken, it is more important to look at the quality of those shots. Since Ryder's paper, expected goals have evolved and there have been several NHL expected goals models put forth which typically rely on machine learning to value a given shot. However, it should be noted that many of these models exist outside the academic sphere \cite{hockeyGraphs,moneyPuck,macdonald2012expected}. Expected goals models are not limited to only the NHL and hockey, as the metric is arguably more popular in European football/Soccer. Thus, there are a host of expected goals models for the sport \cite{green2012assessing,rathke2017examination,hewitt2023machine}.

While expected goals have been accepted as a good indicator of future goal-scoring in the NHL \cite{hockeyGraphs}, the performance indicator often struggles to account for players who are either skilled or unskilled shooters. There have been several proposed adjustments to account for shooter skill such as the model developed by Dawson Sprigings and Asmae Toumi \cite{hockeyGraphs} or the model created by Harry Shomer \cite{shomerTalent}. Both of these approaches resulted in no gains in model performance or very minimal gains. While both of these models provided valuable insight into the problem of shooter skill in the NHL, they both did not account for the skill of the goaltender. They also fail to explore any subsection of skill (i.e. a player may be more skilled when shooting from a certain location or in a certain situation). This is an area of study that this research aims to further the understanding of. 

The approach outlined in this paper accounts for the skill of both the shooter and goaltender, while also accounting for three separate subsections of skill, these subsets being overall skill, locational skill, and situational skill. The assumption here is that by accounting for different subsections of both shooter and goaltender skill, a skill-adjusted expected goals model will have more predictive power than a baseline model that does not account for skill.

The models in this paper are engineered using an architecture similar to Shomer's \cite{shomerTalent}. In this architecture, two expected goals models are stacked such that the second model uses the xG values obtained in the first model to generate features. However, it differs from the aforementioned approach as rather than only using the last two years to account for talent, all available 5v5 shots from 2010-2022 are used to judge a shooter's or goaltender's skill, these shots are then linearly weighted in order of occurrence to better exemplify the player's skill over their entire career. Another distinguishing factor is that this approach creates models for different skill brackets of players and goaltenders. This action was taken as the factors that lead do a goal could be different based on the skill of the individuals involved.

The use of this approach should create a clearer picture of how skill is related to shot outcome, how skill-based expected goals models correlate with a shooter or goaltender's actual performance, and what this means for other types of prediction models in the NHL. This paper should also serve as one of the only comprehensive outlines of an NHL expected goals model in academic literature.

\section{Literature Review}
Early work in the area of expected goals in hockey can be seen in work by Alan Ryder, who proposed a shot quality metric in 2004 \cite{ryder}. Ryder's approach made use of data from the 2002-03 NHL season. Ryder used several different situational factors affecting each shot such as the distance, shot type, and whether the shot was a rebound to determine the likelihood that a given shot would become a goal.

At this same time, in the world of football/soccer. Ensum, Pollard, and Taylor used logistic regression to study the factors that affected the likelihood of a shot becoming a goal during the 2002 World Cup \cite{ensum2005applications}. They found factors that are still used in models today such as distance and angle. 

The work of Ryder \cite{ryder}, as well as Ensum, Pollard, and Taylor \cite{ensum2005applications} laid the foundations for what would become the modern expected goals model by identifying factors that increased the likelihood of scoring.

In 2012, Brian Macdonald presented a paper on expected goals in the NHL at the MIT SLOAN sports analytics conference \cite{macdonald2012expected}. This paper made use of situational factors along with a ridge regression model to create an adjusted plus-minus metric. This metric was then used to measure the positive or negative impact a player had for their team when they were on the ice. For instance, a player would have a positive plus-minus if they had a positive effect on their team's expected goals when they were on the ice. This is a very prominent piece of literature regarding NHL expected goals models as it focuses heavily on how expected goals can be used to evaluate players.

This literature review would be incomplete if it did not acknowledge the findings of Sprigings and Toumi \cite{hockeyGraphs} as well as Shomer \cite{shomerTalent}. Although both of these works took place outside the academic sphere, they still outline approaches for creating skill-adjusted expected goals models using NHL data. Sprigings and Toumi took a simple approach to account for player skill. They did this by including the percentage of a player's shots which resulted in goals as a model feature \cite{hockeyGraphs}. Sprigings and Toumi saw no improvement in their log loss or area under the ROC curve when using the skill-adjusted model. Later, Shomer attempted to iterate on this approach. Shomer argued that the use of shooting percentage is counter-intuitive to the concept of xG. This is due to the fact, that shooting percentage fails to account for the quality of each shot, while the purpose of expected goals is to better understand shot quality. Therefore, they focused on stacking ML models to account for players scoring above or below their expected goals over the last two seasons \cite{shomerTalent}. Shomer saw no improvement in the log loss of their model but saw a very modest improvement in the area under the ROC curve. In both of these instances, the authors came to the conclusion that while skill must play some factor in the likelihood that a shot will result in a goal, the biggest determining factor is the situation in which the shot is arising.

More recently in 2023, Hewitt and Karaku published a work on expected goals models in football/soccer. In this work, the authors proposed a position-adjusted approach to expected goal models which splits a given expected goals model into separate models based on the position of the player, finding that attacking players are better at accumulating expected goals \cite{hewitt2023machine}. The authors also explored player-specific expected goals models that trained on a given player's shots, they found that Lionel Messi was significantly more efficient than the average player.

In the same year as Hewitt and Karaku; Mead, O'Hare, and McMenemy published research that looked at using less traditional model features to improve the capabilities of expected goal models. They experimented with features such as player value and ELO rating, finding that such features could improve the predictive power of models \cite{mead2023expected}.

This paper differentiates itself from these previous works by not only accounting for the skill of the shooter but also the skill of the goaltender. This research aims to account for subsections of shooter and goaltender skill; namely, overall skill, locational skill, and situational skill while also splitting the data set on shooter and goaltender skill to better serve individual skill brackets. Most NHL expected goals models are fenwick-based, meaning they include any shot attempt that resulted in a goal, save, or miss. The approach in this paper differs from previous works as it is shot-based, meaning that only shots that were on-net are included. This was done because it would be incorrect to attribute skill to a goaltender when a shooter fails to hit the net.

\section{Methods}
    \subsection{Data Retrieval and Cleaning}
    Play-by-play data from the 2010-2022 NHL seasons was scraped from the public NHL API using the Python module hockey scraper \cite{hockey-scraper}. Some information about individual players, such as their handedness was scraped with the R package nhlapi \cite{nhlapi}.

    The play-by-play data required cleaning before it could be used in a model. First, all shots that were missing x or y locations were removed from the dataset as well as any shots that had no recorded shot type. Only shots that took place in the shooting team's offensive zone were included in the data. From there the x and y positions for each shot were standardized to the right side of the ice, this allows the model to better identify trends in the positional data. Next, all shots not taken in a five-on-five (5v5) state were removed from the dataset. 5v5 is the default state in which hockey is played, for this reason, it is also the most common.

    After Ryder's initial work on shot quality, he released a paper called ``Product Recall Notice for Shot Quality". This work focused on his findings that there may exist systemic bias in shot data, meaning that certain scorekeepers may be plotting shots closer to the net for the home team, thus introducing biases to shot quality calculations \cite{ryder2007product}. He proposes that using road shot quality is a safer alternative.
    Ryder was not the only person to have concerns with the NHL's shot locations, and in 2013 Shuckers and Curro proposed a method to account for this bias \cite{schuckers2013total}. Using a cumulative distribution function (CDF) they determine the level of bias for each NHL stadium, using this they then adjust the shot locations. Another method was proposed by Ken Krzywicki which adjusted shot distance by subtracting the expected shot distance \cite{krzywicki2009removing}. Both Schuckers and Curro's method as well as Krzywicki's method were employed in this paper. Shuckers and Curro's method was implemented using a Python module called NHLArenaAdjuster \cite{nhlarenaadjuster}.
    
    \subsection{Base Feature Engineering}
        \subsubsection{Extracted Features}
        There are scarce peer-reviewed scientific sources to determine what features to use in an NHL expected goals model. However, there does exist a large amount of work on the topic in the public domain. Websites such as Moneypuck \cite{moneyPuck}, and Evolving Hockey \cite{EvolvingHockey_2021} have information about how their models are comprised available on their websites. Most of the features engineered in the base model were at least inspired by these sources. The features extracted and used in the base xG model can be seen in table \ref{tab:tab1}.

        \begin{table}[H]
            \centering
            \begin{tabular}{|c||p{8.5cm}|}
                \hline
                \textbf{Feature} & \textbf{Description} \\
                \hline
                \hline
                isStrongSide & Is the player’s stick on the side of their body closest to
                the goal.\\
                \hline
                x & The reported x location of the shot, standardized to the right side of the ice.\\
                \hline
                y & The reported y location of the shot, standardized to the right side of the ice.\\
                \hline
                GameTime & The number of seconds that have passed in the game.\\
                \hline
                PeriodTime & The number of seconds that have passed in the period.\\
                \hline
                Distance & The distance the shot was from the net.\\
                \hline
                Angle & The angle between the net and the shooter.\\
                \hline
                ShotType & The type of shot that was taken.\\
                \hline
                GoalDiff & The goal differential between the two teams.\\
                \hline
                LastEvent & The last recorded play-by-play event.\\
                \hline
                LastEventDistance & The distance between the last recorded play-by-play event and the current shot.\\
                \hline
                LastEventZone & The zone the last event recorded took place in relative to the current shooting team.\\
                \hline
                LastEventAngle & The angle between the last recorded play-by-play event and the current shot.\\
                \hline
                LastEventSpeed & The speed between the last recorded play-by-play event and the current shot.\\
                \hline
                TimeSinceLastEvent & The number of seconds that have passed between the last recorded play-by-play event and the current shot.\\
                \hline
                Rebound & Was the shot a rebound? (Was there a shot by the same team in the last 2 seconds?)\\
                \hline
                FlurryCount & If the shot was a rebound, how many rebound shots preceded it?\\
                \hline
                Fastbreak & Was the shot taken on a fastbreak?\\
                \hline
                krzywickiX & The x coordinate when using Krzywicki's venue adjustment.\\
                \hline
                krzywickiY & The y coordinate when using Krzywicki's venue adjustment.\\
                \hline
                krzywickiDist & The distance when using Krzywicki's venue adjustment.\\
                \hline
                SchuckersX & The x coordinate when using Schucker's and Curro's venue adjustment.\\
                \hline
                SchuckersY & The y coordinate when using Schucker's and Curro's venue adjustment.\\
                \hline
                SchuckersDist & The distance from the net when using Schucker's and Curro's venue adjustment.\\
                \hline
                SchuckersAngle & The angle of the shot when using Schucker's and Curro's venue adjustment.\\
                \hline
                Outcome & The outcome of the shot (1 = Goal, 0 = No Goal)\\
                \hline
            \end{tabular}
            \caption{Features that were extracted from the play-by-play data and used in the base xG model.}
            \label{tab:tab1}
        \end{table}

    \subsection{Skill-based Feature Engineering}

    Skill-based features were engineered by looking at all previous shot instances (starting at the previous game) for a given shooter and goaltender. From there the xG value and outcome of each of those previous shots is collected and transformed.

        \subsubsection{Shot Weighting}
        To better capture the current skill level of a given shooter or goalie, skill-based features linearly weighted both xG values and outcomes in order of their occurrence with the most recent shot/outcome having the greatest weight. This approach was taken as it can capture how a shooter or goaltender has played in the past, however, due to the weighting it will be more representative of a player's recent form. Thus, accounting for any change in shooter/goaltender talent over time, such as talent regression in older players and talent growth for younger players. A simple example of this weighting technique can be seen below in Table \ref{tab:tab2}, in this instance shot 1 would be the earliest shot with shot 5 being the most recent.

        \begin{table}[H]
            \resizebox{\textwidth}{!}{
            \centering
            \begin{tabular}{|c|c|c|c||c|c|c|}
            \hline
            Shot Number & Shot Weight & Original xG Value & Original Outcome & \textbf{Weighted xG Value} & \textbf{Weighted Outcome} \\
            \hline
            \hline
            1 & 0.2 & 0.2 & 0 & 0.04 & 0\\
            2 & 0.4 & 0.03 & 1 & 0.012 & 0.4\\
            3 & 0.6 & 0.05 & 0 & 0.03 & 0\\
            4 & 0.8 & 0.4 & 1 & 0.32 & 0.8\\
            5 & 1.0 & 0.06 & 0 & 0.06 & 0\\
            \hline
            \end{tabular}%
            }
            \caption{An example of how xG values and outcomes are weighted in skill-based feature engineering.}
            \label{tab:tab2}
        \end{table}

        \subsubsection{Shot Binning}
        In order to further differentiate this work from other skill-adjusted approaches, a subsection of features were engineered while only looking at shots that took place in the same bin. Shots were binned using a method proposed by Shuckers and Curro \cite{schuckers2013total}. This method involves dividing the offensive zone between the blue line and the goal line into nine equal bins. The method includes two extra bins, one that represents shot attempts that took place outside of the offensive zone (either in the neutral zone or defensive zone of the team who took the shot) and one that represents shot attempts that took place below the goal line. As this paper only deals with shots that took place in the offensive zone the former extra bin will be disregarded. This binned approach was used as it is a possibility certain shooters are more proficient when shooting from certain locations on the ice. At the same time, it is also possible that certain goaltenders may be stronger when facing shots from certain locations on the ice. Therefore, the inclusion of both binned features and features that encapsulate all shots, allows the model to derive a shooter's or goaltender's overall skill along with their general skill regarding this location on the ice.

        \subsubsection{Gower Distance}
        Gower distance is a measurement that accounts for the similarity between two sets of variables, it is capable of accounting for both numerical and categorical values \cite{gower1971general}. In this paper, for each given shot the Gower distance between that shot and all previous shots taken by a given shooter or faced by a goaltender was computed using only a subset of all the baseline features. This subset is outlined in Table \ref{tab:tab3} and was used to improve computational efficiency and reduce redundancy (i.e. there is no need to compare 3 separate x coordinates). These distances were then inversely normalized such that the value with the smallest distance would have a value of one. The xG and outcome for each shot were then multiplied by each shot's gower distance. Lastly, the shots were then sorted by occurrence and linearly weighted as was discussed in the previous section. This allows the model to determine the situational skill of a given player or goaltender. 

        \begin{table}[H]
            \centering
            \begin{tabular}{|c||p{8.5cm}|}
                \hline
                \textbf{Feature} & \textbf{Description} \\
                \hline
                \hline
                isStrongSide & Is the player's stick on the side of their body closest to the goal.\\
                \hline
                LastEvent & The last recorded play-by-play event.\\
                \hline
                ShotType & The type of shot that was taken.\\
                \hline
                SchuckersX & The x coordinate when using Schucker's and Curro's venue adjustment.\\
                \hline
                SchuckersY & The y coordinate when using Schucker's and Curro's venue adjustment.\\
                \hline
                SchuckersDist & The distance from the net when using Schucker's and Curro's venue adjustment.\\
                \hline
                SchuckersAngle & The angle of the shot when using Schucker's and Curro's venue adjustment.\\
                \hline
                Rebound & Was the shot a rebound?\\
                \hline
                Fastbreak & Was the shot taken on a fastbreak?\\
                \hline
            \end{tabular}
            \caption{Features that were used to calculate gower distance.}
            \label{tab:tab3}
        \end{table}

        \subsubsection{Extracted Features}
        Several skill-based features were engineered for the skill-adjusted model. All features had three separate instances engineered, one that looked at all available shots, one that looked only at shots that took place in a given bin, and one that used Gower distance in order to weigh shots. This results in creating three separate instances of skill evaluation for players and goaltenders. A total skill, a locational skill, and a situational skill.
        
        The first type of feature engineered was an ``above-expected" feature. This involved calculating how many goals the shooter had scored above expected, and how many goals the goaltender had saved above expected. These formulas can respectively be seen in equations \ref{eq1} and \ref{eq2}. 

        \begin{equation}
        \resizebox{0.9\hsize}{!}
        {\text{Shooter Goals Above Expected} = \text{Total Goals Scored Weighted} - \text{Cumulative Weighted xG}}
        \label{eq1}
        \end{equation}

        \begin{equation}
        \resizebox{0.9\hsize}{!}
        {\text{Goalie Goals Saved Above Expected} = \text{Cumulative Weighted xG} - \text{Total Goals Scored Weighted}}
        \label{eq2}
        \end{equation}

        The second feature engineered was a talent feature. This is the same feature Shomer outlined in their previous work \cite{shomerTalent}, however, in this instance, all available shots are being used and weighted. This feature is a ratio, in the case of the shooter, it is the ratio of goals scored over xG. For the goaltender, this ratio was reversed to be the ratio of xG against over goals scored against. In cases where a shooter or goaltender had no available shot data, their talent was set to zero. The player talent calculation can be seen in equation \ref{eq3}, with the goalie talent calculation visible in equation \ref{eq4}.
        
        \begin{equation}
        \text{Shooter Talent} = \frac{\text{Cumulative Weighted Goals Scored}}{\text{Cumulative Weighted xG}}
        \label{eq3}
        \end{equation}
        
        \begin{equation}
        \text{Goalie Talent} = \frac{\text{Cumulative Weighted xG Against}}{\text{Cumulative Weighted Goals Scored Against}}
        \label{eq4}
        \end{equation}

        Using the features engineered so far a ``True" Talent and above-expected feature was calculated for both the shooter and goaltender. These equations involved adding the above-expected or talent features for the three separate skill types outlined (total skill, locational skill, and situational skill) together. This can be seen in equations \ref{eq5} and \ref{eq6}.
        
        \begin{equation}
        \resizebox{0.9\hsize}{!}
        {\text{True Player Talent} = \text{Total Player Talent } + \text{ Locational Player Talent } + \text{ Situational Player Talent} }
        \label{eq5}
        \end{equation}

        \begin{equation}
        \resizebox{0.9\hsize}{!}
        {\text{True Player Above-Expected} = \text{Total Player Above-Expected } + \text{ Locational Player Above-Expected } + \text{ Situational Player Above-Expected} }
        \label{eq6}
        \end{equation}
        
        When all calculations were finished the skill-adjusted the features seen in table \ref{tab:tab4} were used to comprise the skill-adjusted model.

        \begin{table}[H]
            \resizebox{\textwidth}{!}{
            \centering
            \begin{tabular}{|c|c|}
                \hline
                \textbf{Feature} & \textbf{Explanation}\\
                \hline
                \hline
                True Goals Scored Above-Expected Shooter & The sum of the above-expected features for the shooter.\\
                \hline
                True Talent Shooter & The sum of the talent calculations for the shooter.\\
                \hline
                True Goals Saved Above-Expected Goaltender & The sum of the above-expected features for the goaltender.\\
                \hline
                True Talent Goaltender & The sum of the talent calculations for the goaltender.\\
                \hline
                xG & The xG value assigned to the shot by the baseline model.\\
                \hline
            \end{tabular}
            }%
            \caption{The features that comprised the skill-adjusted model.}
            
            \label{tab:tab4}
        \end{table}

    \subsection{Model Creation \& Training}
    All models were built using light gradient boosting in Python. These models had their hyperparameters tuned using Optuna \cite{optuna_2019}, a Python module that employs Bayesian optimization and cross-validation to find the ideal hyperparameters for a given model. Past studies have found that Optuna has a good trade-off between runtime and predictive performance \cite{shekhar2021comparative}. Therefore, it was employed to tune parameters for both the base and skill-based models.

    Models were created based on skill, which was defined as Cumulative Goals divided by Cumulative xG for shooters and Cumulative xG Against divided by Cumulative Goals Against for goaltenders. The data was split based on skill brackets (low-skill, medium-skill, and high-skill). These ranges were defined by percentiles for shooters and goaltenders, these ranges were $p \leq 0.5$, $0.5 < p \leq 0.75$, and $p > 0.75$. This means that technically six separate models were defined, two for each skill bracket, one baseline model, and one skill-adjusted model. This was done as it is possible that the features that were most important in determining if a shot became a goal for a low-skill shooter would not be as impactful on a high-skill shooter and vice versa. This model setup ensured that when predicting shot outcomes the model was trained on shots involving players of a similar skill level.

    The process of creating the models was as follows. First, an initial base model was created using the base feature set. This model trained and tuned its hyper-parameters using data from the 2010 through 2020 seasons. This model then predicted shot outcomes for the 2021 and 2022 NHL seasons. In order to construct a skill-based model, xG values for all shots would have to be available. Therefore, the predicted shot outcomes for 2021 and 2022 were joined with predictions for all shots for the seasons of 2010 through 2020. This was attained using cross-validation of the training set, namely through the cross\_val\_predict method from the scikit-learn module \cite{sklearn_api}.

    Using the xG values obtained from the base 2010-2022 model the skill-based features were engineered. These features were only engineered for shots from 2012 to 2022. This was done to ensure there was a sufficient sample of shots for veteran players in the early seasons (i.e. 2012). The skill-adjusted feature set was then used to construct a skill-adjusted model that trained and tuned its hyper-parameters using data from the 2012 through 2020 seasons. Following this, the skill-adjusted model predicted the held-out seasons of 2021 and 2022. 

    Lastly, another base model was constructed, this time training and tuning its hyper-parameters using data from the 2012 to 2020 season, while testing on the 2021 and 2022 seasons. This is the base model that will be compared to the skill-adjusted model. This ensures that both the skill-adjusted and base models have an equal number of training and testing instances.

    \subsection{Model Comparison}
    The goal of creating both a skill-adjusted model and a base model is to provide a comparison between these two approaches and determine if the outlined skill-adjusted approach accurately accounts for shooter and goaltender skill. These models were only compared on the results from predicting the held-out seasons of 2021 and 2022, this was done to ensure there was no leak of information to the skill-adjusted model in the process of cross-validation. These models were bench-marked using three metrics. These metrics were the area under the ROC curve, the log loss, and the Brier score.

    To better understand the difference between low-skill, mid-skill, and high-skill players, the normalized feature importance based on the gain for each model is used. This will allow comparison between factors affecting the likelihood of a given shot becoming a goal based on the skill of the players involved.

\section{Results}

The results seen in Table 5 show how the baseline and skill-adjusted models compare to each other. In all skill ranges for all metrics, the skill-adjusted model is a better predictor of shot outcomes than the baseline model. There are differing magnitudes of difference among the models when broken down into skill levels and individual metrics.

\begin{table}[H]
    \centering
    \begin{tabular}{|c||c|c|c|}
        \hline
        \multicolumn{4}{|l|}{\textbf{Baseline Model}} \\
        \hline
        \hline
        \textbf{Percentile ($p$)} & \textbf{Log Loss} & \textbf{AUC} & \textbf{Brier Score}\\
        \hline
        $p > 0.75$ & 0.2982 & 0.7238 & 0.0849 \\
        $0.5 < p \leq 0.75$ & 0.2831 & 0.7424 & 0.0809 \\
        $p \leq 0.5$ & 0.2126 & 0.7531 & 0.0567 \\
        \hline
        \hline
        \multicolumn{4}{|l|}{\textbf{Skill Adjusted Model}} \\
        \hline
        \hline
        \textbf{Percentile ($p$)} & \textbf{Log Loss} & \textbf{AUC} & \textbf{Brier Score}\\
        \hline
        $p > 0.75$ & 0.2844 & 0.7616 & 0.0816 \\
        $0.5 < p \leq 0.75$ & 0.2792 & 0.7519 & 0.0798 \\
        $p \leq 0.5$ & 0.2100 & 0.7630 & 0.0560 \\
        \hline
    \end{tabular}
    \caption{A comparison of metrics achieved by the baseline and skill-adjusted models.}
    \label{tab:tab5}
\end{table}

\begin{figure}[H]
    \centering
    \includegraphics[width=1\textwidth]{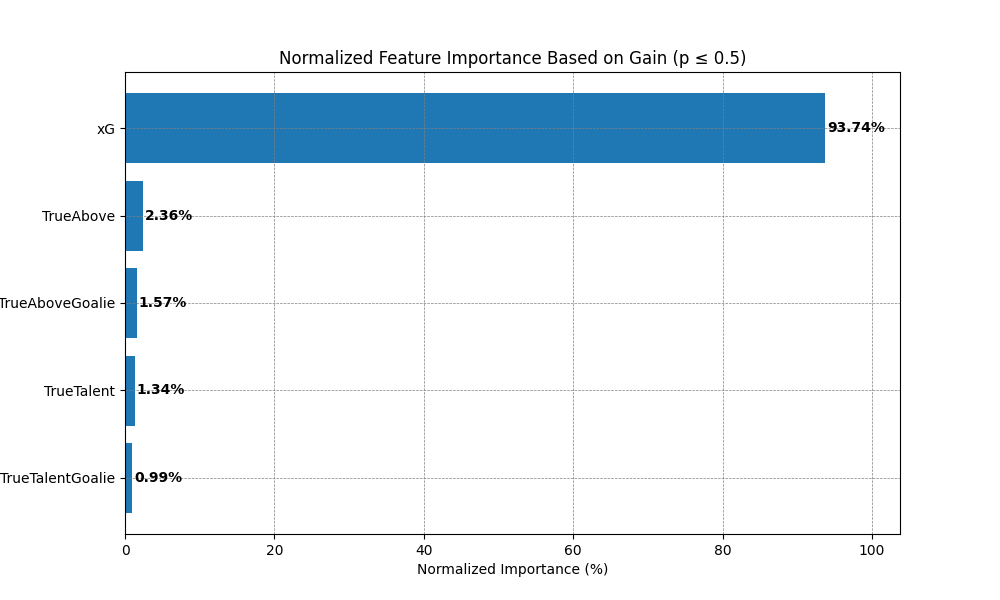}
    \caption{Feature importance based on gain for the low-skill model.}
    \label{fig:fig1}
\end{figure}

\begin{figure}[H]
    \centering
    \includegraphics[width=1\textwidth]{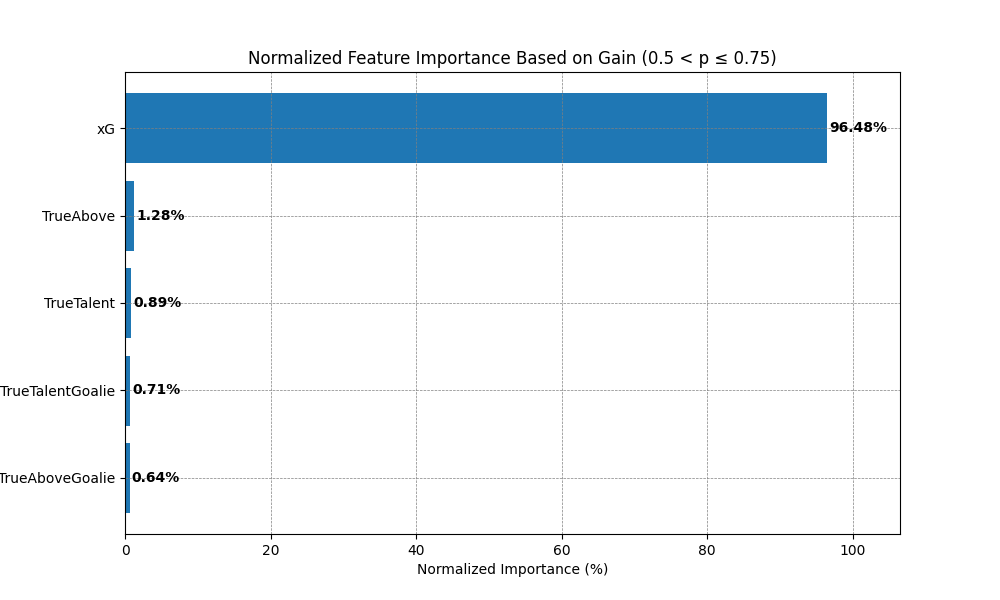}
    \caption{Feature importance based on gain for the mid-skill model.}
    \label{fig:fig2}
\end{figure}

\begin{figure}[H]
    \centering
    \includegraphics[width=1\textwidth]{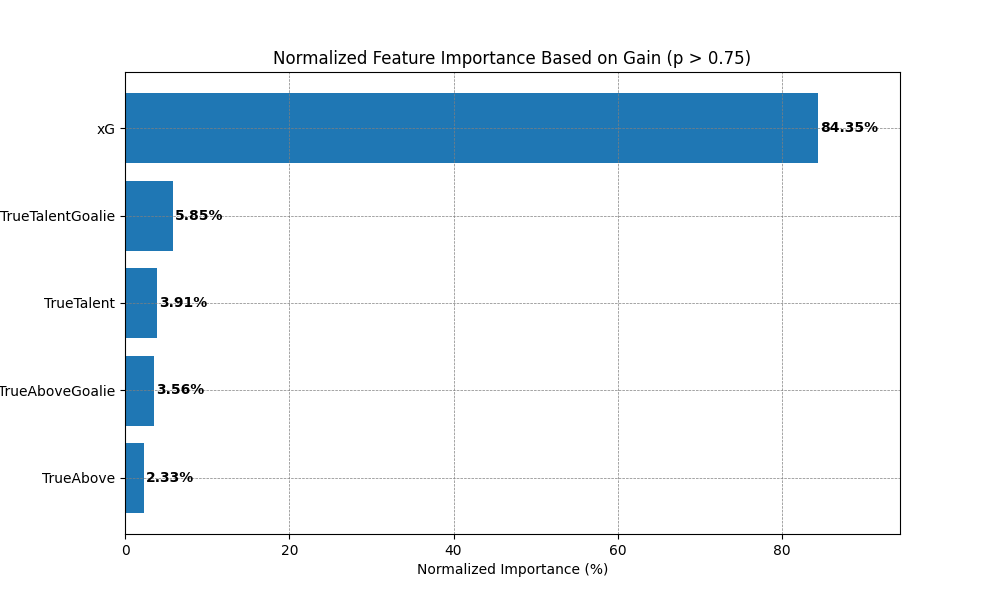}
    \caption{Feature importance based on gain for the high-skill model.}
    \label{fig:fig3}
\end{figure}

Figures \ref{fig:fig1}, \ref{fig:fig2}, and \ref{fig:fig3} show the feature importance based on gain for each skill-adjusted model. In all three models, the xG value from the baseline model is the most impactful feature by a very large margin. However, both low-skill and high-skill models seem to make use of the skill-based features, with the high-skill model making the most use of the engineered skill-based features. 

\section{Discussion}

When looking at the results found in Table \ref{tab:tab5}, it is possible to discern several things. The skill-adjusted model outperforms the baseline model in every outlined metric. Although these performance increases are not monumental, the increases are significantly greater than was seen in previous works \cite{hockeyGraphs,shomerTalent}. This could be for several different reasons, such as the years the models used for training and testing or differences in the basic feature engineering between models. However, the approach outlined in this paper attempted to isolate shooter and goaltender skill into locational, situational, and overall skill. As this is a new approach in expected goals models it is fair to make the assumption that the isolation of these factors had a role in the performance increases seen. The differences in log loss, AUC, and Brier score between the skill-adjusted and baseline models remain small with the performance increases being seen lying in the range of 1-5\%. However, it should be noted that the high-skill model is consistently on the high side of this range, giving the impression that the skill-adjusted approach is more suited to high-skill shooters and goaltenders.

When looking at the feature importance in figures \ref{fig:fig1}, \ref{fig:fig2}, and \ref{fig:fig3}. It is evident that skill-based features are being utilized more in the high-skilled model. All three models are mainly affected by the xG value from the baseline model however, the high-skilled model is also affected by both the skill of the shooter and the goaltender. The low-skill model is slightly affected by the skill of the shooter. Lastly, the mid-skill model is not significantly affected by the skill of the shooter or the goaltender.

The fact that skill-adjusted xG models have more predictive power than baseline models has several implications for real-world hockey situations. This implies that the skill of both the shooter and goaltender does indeed have an impact on the likelihood that a given shot will become a goal even if that impact is not as large as what may be perceived by fans, coaches, and players. This allows us to add another layer of depth to Alan Ryder's comments that not all shots are created equal, as not only are shot outcomes influenced by location and situation but also by the skill of the players involved. Often xG models are used in conjunction with other performance indicators to predict the outcome of a given game. Given that the skill-adjusted model can improve the performance of the baseline model it is possible that including skill-adjusted xG in a pre-game prediction model could lead to improvements in the predictive power of said models.

While it is hard to draw implications from these findings for goaltenders; the findings have implications as to what makes a goal-scorer in the NHL. While skill does have an impact on the outcome of a shot, ultimately a goal-scorer is defined by their ability to put themselves in quality scoring positions. This indicates that a shooter's intelligence and the way they perceive the game is paramount to their success as a goal-scorer. This is very difficult to quantify, however, metrics such as expected goals can help us better understand a player's decision-making capabilities. 

This work can also provide new insight into the ever-growing subsection of sports literature that studies the randomness of outcomes. Lopez et al. previously found that game outcomes in the NHL are more random when compared to other major leagues in North America \cite{lopez2018often}. Similar results are also seen in the work of Gilbert and Wells \cite{gilbert2019ludometrics}. As has already been stated, this paper has found that implementing a skill-adjusted expected goals model can yield better performance than a baseline model, however, it still shows that skill is a small part of the equation and that in general, the quality of the shot is the most important factor when determining if a shot becomes a goal. It is fair to assume that luck also plays a part in shot outcomes otherwise the predictive power of a skill-adjusted model would be far higher. This can support the claim that luck has a significant effect on sports as a whole, not only does luck affect outcomes in the NHL but it also affects whether shots become goals.

There are several potential shortcomings when looking at this work. For instance, due to the nature of the event data for the NHL, it is impossible to know the positional coordinates of anyone but the shooter. However, other players on the ice can potentially have a significant impact on the likelihood of a goal. If a goaltender is unable to see the shooter due to an opposing player who is ``screening" them (standing in front of the goaltender such that they cannot see an oncoming shot), the shot could have an increased likelihood of becoming a goal. To combat this problem full player tracking data would have to be readily available to the public, however, at the time of writing it is not. Internally, NHL teams have access to this data and may also consult with one of the major private organizations in sports analytics such as SportsLogic. 

Another potential shortcoming is how the baseline model was constructed. The model was constructed using features highlighted in other public models but was not constructed to be the highest-performing public model available. This by itself would require its own publication with a heavy emphasis on data cleaning, feature engineering, model selection, and hyperparameter tuning. Therefore, it is fair to assume there is some performance left on the table. 

This study only focused on 5v5 performance which can be viewed as a shortcoming as this is only a subsection of the game. In future publications, it would be beneficial to attempt to include all game states in the model and determine how skill affects shooters and goaltenders in different game states (i.e. 5v5, 5v4, etc.). It would also be worth testing how this approach to including skill in xG models applies to football/soccer. This would allow a comparison between the effects of shooter and goaltender skill in ice hockey and football/soccer. In future work, it would also be worth taking a causal approach to the skill of shooters and goaltenders. This should allow us to better understand the relationship between skill and shot outcomes.

\section{Conclusion}

The purpose of this research was to provide an NHL expected goals model capable of accounting for both the shooter's and the goaltender's skill. The findings here show that while accounting for the overall, locational, and situational skill of both shooters and goaltenders, there are performance increases across all metrics used. However, these increases are in many cases minimal. Therefore, it is fair to conclude that while skill does play a part in whether or not a given shot becomes a goal it is not the sole determinate. It also shows that oftentimes players become elite goal scorers due to their ability to put themselves in goalscoring situations.

In the future, it would be beneficial to determine how this method of accounting for shooter and goaltender skill affects predictions made at other game states such as the power play. It would also be interesting to see how this research can be applied to European football/soccer.

\printbibliography

\end{document}